# AB-Net Method of Protection from Projectiles (city, military base, battle-front, etc.)

*Alexander Bolonkin*

C&R, 1310 Avenue R, #F-6, Brooklyn, NY 11229, USA

T/F 718-339-4563, aBolonkin@juno.com, or aBolonkin@gmail.com http://Bolonkin.narod.ru

## Abstract

The author suggests a low cost special AB-Net from artificial fiber, which may protect cities and important objects from rockets, artillery and mortar shells, projectiles, bullets, and strategic weapons. The idea is as follows: The offered AB-Net joins an incoming projectile to a small braking parachute and this incoming projectile loses speed by air braking after a drag distance of ~50 – 150 meters. A following interception net after the first may serve to collect the slowed projectiles and their fragments or bomblets so that they do not reach the aimpoint. The author offers the design of AB-Net, a developed theory of snagging with a small braking parachute by AB-Net; and sample computations. These nets may be used for defense of a town, city, military base, battle-front line, road (from terrorists), or any important objects or installations (for example nuclear electric station, government buildings, etc.). Computed projects are: Net to counter small rockets (for example, from Qassam), net to counter artillery projectile (caliber 76 mm), net to counter bullets (caliber 7.6 mm).

The offered method is cheaper by thousands of times than protection of a city by current anti-rocket systems. Discussion and results are at the end of the article.

**Key words:** Protection from missile and projectile weapons, mortar, rocket, AB-Net, Qassam defense, incoming defense, armor.

*Note*: Some background material in this article is gathered from Wikipedia under the Creative Commons license.

## Introduction

**Review of current situation and methods of protection**.

*Protection of cities, bases, and important objects*. One important problem in small countries with hostile borders (or larger countries with leaky borders) is protection of abroad military bases, domestic targets from small rockets, missiles, mortar shells and terrorists. For well over a hundred years there has been no satisfactory solution for this, which is why such weapons are favorites of guerilla groups. Israel, for example, has villages (Alumim, and dozens of others) towns (Sderot, Netivot and Qiryat Shemona), cities such as Ashqelon and numerous installations near unfriendly borders. For example, Sderot lies one kilometer from the Gaza Strip and the town of Beit Hanoun. Since the beginning of the Second Intifada in October 2000, Sederot has been under constant rocket fire from Qassam rockets launched by various armed factions. Despite the imperfect aim of these homemade projectiles, they have caused deaths and injuries, as well as significant damage to homes and property, psychological distress and emigration from the city. Real estate values have fallen by about half. The Israeli government has installed a "Red Dawn" alarm system to warn citizens of impending rocket attacks, although its effectiveness has been questioned. Thousands of Qassam rockets have been launched since Israel's disengagement from the Gaza Strip in September 2005, which essentially has killed popular support for any further withdrawals, particularly from West Bank areas near the heart of the country.

In May 2007, a significant increase in shelling from Gaza prompted the temporary evacuation of thousands of residents. By November 23, 2007, 6,311 rockets had fallen on the city. The Israeli newspaper Yediot Aharonot reported that during the summer of 2007, 3,000 of the city's 22,000 residents (comprised mostly of the city's key upper and middle class residents, the heart of the economy, those most able to move,) had already left for other areas, out of Qassam rocket range.

The total number of Qassam rockets launched exceeded 1000 by June 9, 2006. During the year 2006 alone, 1000+ rockets were launched. Tons of explosives have been intercepted at the Egyptian border; the uninterrupted shipments must be greater still, and the cumulative detonation yield has easily been in the tens of tons.

A rocket once fell into the electricity station in Ashkelon and caused electricity shortages in several areas, other time a rocket-similar to Qassam - fell inside an army base and injured more than 70 Israeli soldiers. The Ashkelon strike in particular was troubling as it added (by its radius) another 250,000 Israelis to the potential target list requiring defenses to be paid for, active or passive.

Some military bases of the USA in Afghanistan and Iraq (or in various parts of Asia and Latin America) are in the same situation. Any security consultant working to protect valuable installations in the more volatile corners of Africa, Latin America or Asia will recognize the dangers in the scenarios listed above. Rockets, and remotely triggered mortars, are man-portable and can be smuggled in, can be covertly emplaced and remotely fired with no appreciable warning, and endanger billions in investment with mere thousands in expenses. In Gaza, bonuses are allegedly given to impoverished children to retrieve the launchers, to reduce the expenses of replacing both rounds and launchers.

The Qassam rockets are tens (hundreds) times cheaper than complex electronic-guided anti-rockets and their delicate support system. The mortar shells are cheaper by hundreds of times. Attempted defense from them by conventional anti-rocket system may ruin any rich country, and in the end not work anyway, because the enemy always has the option of using large salvos, to probe the point where the system collapses trying to defeat X+1 simultaneous launches.

**Protection by Armor.**

At present times for defense of small objects armor is employed. For defense of large objects (e.g., a city) from a nuclear warhead the very complex and very expensive anti-rocket systems under development may be used (missile interceptors). Obviously these would be economically infeasible to use against a conventional barrage.

*Review of current armor and defense system*s. Military vehicles are commonly **armored** to withstand the impact of shrapnel, bullets, missiles, or shells, protecting the personnel inside from enemy fire. Such vehicles include tanks, aircraft, and ships.

Civilian vehicles may also be armored. These vehicles include cars used by reporters, officials and others in conflict zones or where violent crime is common, and presidential limousines. Armored cars are also routinely used by security firms to carry money or valuables to reduce the risk of highway robbery or the hijacking of the cargo.

Armor may also be used in vehicles from threats other than deliberate attack. Some spacecraft are equipped with specialised armor to protect them against impacts from tiny meteors or fragments of space junk. Even normal civilian aircraft may carry armor in the form of debris containment walls built into the casing of their gas turbines to prevent injuries or airframe damage should the compressor/turbine wheel disintegrate.

The design and purpose of the vehicle determines the amount of armor plating carried, as the plating is often very heavy and excessive amounts of armor restrict mobility.

Vehicle armor is sometimes improvised in the midst of an armed conflict. In World War II, U.S. tank crews welded spare strips of tank track to the hulls of their Sherman, Grant, and Stuart tanks. In the Vietnam War, U.S. "gun trucks" were armored with sandbags and locally fabricated steel armor plate. More recently, U.S. troops in Iraq armored Humvees and various military transport vehicles with scrap materials: This came to be known as "haji" armor by Iraqis and "hillbilly" armor by the Americans.

For increasing vehicle protection the space armor, composed armor, active protection and other methods.

**Spaced armor**. Armor with two or more plates spaced a distance apart, called spaced armor, when sloped reduces the penetrating power of bullets and solid shot as after penetrating each plate they tend to tumble, deflect, deform, or disintegrate, when not sloped reduces the protection offered by the armor, and detonates explosive projectiles before they reach the inner plates. It has been in use since the First World War, where it was used on the Schneider CA1 and St. Chamond tanks. Many early-WWII German tanks had spaced armor in the form of armored skirts, to make their thinner side armor more effective against anti-tank fire.

**Composite Armor**. Composite armor is armor consisting of layers of two or more materials with significantly different chemical properties; steel and ceramics are the most common types of material in composite armor. Composite armor was initially developed in the 1940s, although it did not enter service until much later and the early examples are often ignored in the face of newer armor such as Clobham armor. Composite armor's effectiveness depends on its composition and may be effective against kinetic energy penetrators as well as shaped charge munitions; heavy metals are sometimes included specifically for protection from kinetic energy penetrators.

**Plastic armor**, called **plastic protection** in the United States, was a type of vehicle armor originally developed for merchant ships by the British Admiralty in 1940. The original composition was described as 50% clean granite of half-inch size, 43% of limestone mineral, and 7% of bitumen. It was typically applied in a layer two inches thick and backed by half an inch of steel.

Plastic armor was highly effective at stopping armor piercing bullets because the hard granite particles would turn the bullet which would then lodge between plastic armor and the steel backing plate. Plastic armor could be applied by pouring it into a cavity formed by the steel backing plate and a temporary wooden frame.

The armor was cheap and easy to install on ships, and the skills and equipment for installation came from the under-utilized road building industry.

Once installed on ships, plastic armor proved highly effective, when applied in sufficient thickness. Many anti aircraft guns such as the Oerlikon were fitted with only very thin plastic shields, which served mainly to improve the morale of the gunner. By some measures, it was as good as plate steel, and was widely adopted by allied ships. In the United States, some 3,000 merchant ships and 1,000 other ships were equipped with it, and in Britain and the Commonwealth some 7,000 ships were fitted.

An **active protection system**, or **APS**, protects a tank or other armored fighting vehicle from incoming fire *before* it hits the vehicle's armor. There are two general categories: *soft kill* systems, which use jamming or decoys to confuse a missile's guidance system, and *hard kill* systems, which attempt to detect and destroy incoming projectiles.

Soft-kill systems were (unsuccessfully) deployed by Iraq in the Gulf War. Iraqi tanks were fitted with strobe lights that masqueraded as the guidance beacon on the back of a TOW missile. The multinational force was aware of their use and adjusted the frequency of their guidance systems so they would not be confused. A soft-kill system currently in service is the Russian Shtora, deployed on Russian and Ukrainian tanks.

Hard-kill systems are activated when a millimetre-wavelength radar or other sensor detects an incoming projectile. In considerably less than a second, they launch a counter-projectile in an attempt to physically damage or destroy the incoming round. Examples include the TROPHY and Iron Fist from Israel and the Russian Drozd and Arena.

Attempts to use aircraft-mounted flak cannon as such an APS against anti-aircraft missiles proved ineffective, Anti-aircraft missiles are designed for effectiveness in a near-miss shot, making APS inefficient and unreliable. Among the effective countermeasures for aircraft are ECM, flares or anti-radar chaff.

DARPA is presently developing the High Energy Liquid Laser Area Defense System, which is planned to be capable of knocking out missiles, and may be used to actively defend future combat aircraft.

Warships have been equipped with similar systems (more frequently known as Close-In Weapon Systems, CIWS), which use small- to medium-caliber (12.7-76mm) guns and guided missiles to destroy inbound missiles and cannon shells. Examples include the US Phalanx CIWS, Dutch Goalkeeper, Russian Kashtan, joint USA/German Rolling Airframe Missile, British Sea Wolf, Chinese Type 730 and Turkish Sea Zenith.

*Shells* can also be divided into three configurations: bursting, base ejection or nose ejection. The latter is sometimes called the shrapnel configuration. The most modern is base ejection, which was introduced in World War I. Both base and nose ejection are almost always used with airburst fuzes. Bursting shells use various types of fuze depending on the nature of the payload and the tactical need at the time.

## Descriptions and Innovations

The main idea of the offered method is using the atmospheric friction gradient as armor against incoming hostile bodies. It is well-known the air produces a drag effect on moving objects, especially on a body which moves with high speed near sea level. For rockets and projectiles this added drag is mission-harmful because it signficantly decreases the projectile distance (range) and piercing force of projectiles, shells and bullets. At distances of ~ 500 – 1000m, as the projectile speed decreases aproximately by two times, the theoretical distance (in airless space) decreases by 2 – 4 times. Artillery designers try to decrease air drag by various clever techniques. Artillery designers model sharp forms for a projectile, specifically to decrease the cross-section area of said projectile.

Our purpose is precisely the opposite – maximally increasing the drag of ENEMY projectiles, decreasing their range and piercing force, changing their trajectory and not allowing the projectile to reach their (defended) aimpoint. The traditional way to affect this to armor the aimpoint. It is possible (in limited cases) when the target is a tank, car or ship. But it is impossible when the target is a city, military base, nuclear electric station or very large government building. For defending these targets we know one method – anti-rockets, ballistic missile defense (BMD). But their design needs very expensive R&D over many years, high technology industry, expensive production facilities and will probably be too expensive to test fully and therefore unlikely to work properly the first few times it is used! The cost of this BMD in many times more expensive than the cost of the intercepted projectile. It is not a practical solution for small or poor countries.

The suggested system, by contrast is cheap and can affordably protect a large territory.

There is a well-known method to increase the air drag – a parachute. The parachute is used for braking a parachutist, or braking an entire aircraft in a runway landing. The main problem in the offered system sounds impossible at first hearing – how to join and mount the parachute to the enemy projectile in just a moment of contact? The invented net for it is described below.

Two interactions may exist between a flying body and a net in the moment of impact (Fig.1):
1. If body mass and speed is small (<50 m/s) and an elastic net is strong, the body fully loses its velocity and is stopped by the net (Fig.1a). An example might be a badminton ball. The energy (speed) of the body is high. The body then tears the net and continues its' flight (Fig.1b). In this case (using special net and design) we can join to the projectile a small braking parachute. The mechanism of joining is simply that of being snagged in a resilent pocket whose sudden jerking deploys a retro-parachute. The projectile quickly dissipates the high speed within a distance of ~50 – 150m. The next elastic net located after the parachute-affixing net (Fig.1b, notation 9) can capture and collect the projectile and its bomblets, if any have dispersed. (option).

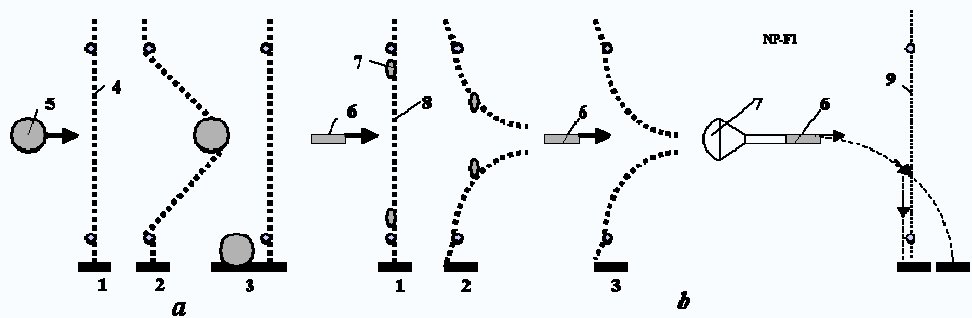

**Fig.1.** Two models of body impact in net: (*a*) Weak impact (the body speed is small, less than 20 – 50 m/s); (*b*) Strong impact (the body speed is high, more than 50 m/s). *Notations*: 1 – position of body and net before impact; 2 – position at impact; 3 - position after impact; 4 – brake net; 5 – ball; 6 – projectile (bullet); 7 – small braking parachute; 8 –parachute net; 9 – capture net.

The design of parachute net is shown in Fig.2. That has a big strong net 10. The thin light sub-net 11 made from high-strength artificial fiber and located in a cell of the big net. The net 11 contains the packed small light parachute 12 and a copy of this is located into every cell of the big net. Internal subnets are weakly joined to the main big net. When an incoming projectile penetrates into a given subnet, it catches on the strong light elastic subnet together with its' rapidly deploying small light parachute and is snagged, ripping it off and triggering deployment. The parachute opens and brakes the projectile. The big net now has a small hole. The cell of subnet can have 1, 2, or 4 parachutes as it is shown in fig. 2b.

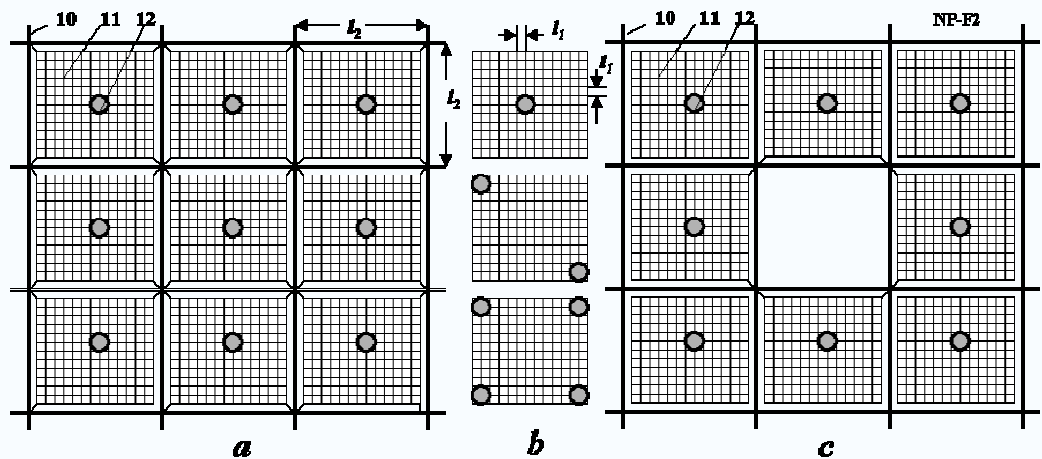

**Fig.2.** Design of parachute net. (*a*) Brake net before impact; (*b*) Location of small braking parachutes (1, 2, 4) in small section (cell) of the braking subnet; (*c*) braking net after projectile impact. *Notations*: 10 – strong cables of braking net; 11 – small section (cell) of braking subnet weak connected to strong cable; 12 – small braking parachute in compact form; $l_1$ is stepping distance of fibers (cell size) in subnet; $l_2$ is stepping distance of fibers (the cell size) in the main net.

The impact of projectile into the subnet is shown in Fig.3. The cell size of the subnet must be less than the diameter of any likely projectile. (This obviously also states that there are different mesh sizes of AB-Nets for different incoming ammunition—one layer of AB-Net might pass bullets but stop small mortars, another might shield for yet larger incoming aerial bombs, etc.) The maximum figure for this cell size is one half of the projectile diameter. If strength of subnet cables is not enough, the possible places 15 of subnet cable break-off-points of parachute subnet is shown in Fig.3a.

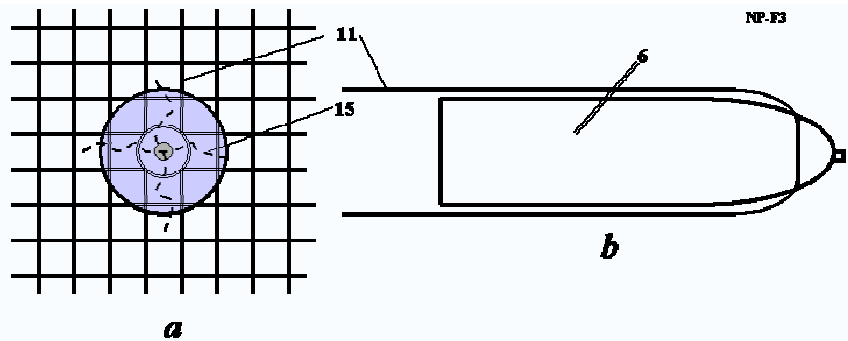

**Fig.3.** Impact of projectile in parachute net: (*a*) Forward view; (*b*) Side view. *Notations*: 11 –parachute subnet; 15 - possible places of subnet cable break-off-point of parachute subnet.
The projectile can snap the break-off-point of between 2, 4 or 8 subcables. The exact figure depends upon the relation between subnet stepping distance (cell size) $l_1$ and the diameter and orientation of the incoming projectile.

The air drag of the parachute (and stress on it) may be very high in the first moment of opening. Although hard on the missile, this is also hard on the parachute! A better design is a slow opening parachute as shown in fig.4. The first moment the parachute has demi-form (fig.4a), when the speed and drag decreases, the parachute opens to full form (fig.4b).

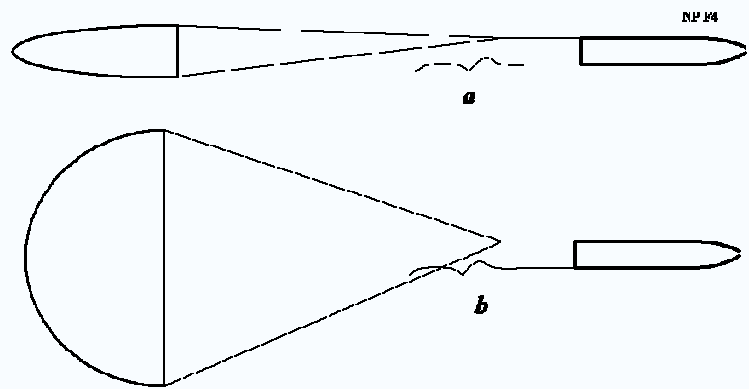

**Fig.4.** Better work of braking projectile parachute: (*a*) Parachute in demi-form (semi-braking); (*b*) Parachute in full form (full braking when projectile speed decreases).

The protection (covering) of a big city by offered AB-Net is shown in Fig,5. Inflatable sectionated stacked balloon masts (imagine a stack of doughnut balloons) 21 (Fig.5a) (about 10 - 20m of diameter and height 300 -500m) are installed over a distance of about 0,5 – 1 km and supported by bracing cables. The masts support the parachute net 22. The capture net 23 is located below this parachute net by a margin of about ~150 – 200m. The debris net 24 is located below the capture net 23 by a margin of about 50m.
  The AB-Net protection system works the following method. The upper parachute net 22 joins the braking parachutes to rockets and projectiles. In a braking distance of 150 – 200m, they decrease projectile speed to the range of 25 – 40 m/s. The lower capture net 23 captures the projectile. If it explodes,, the lower debris net 24 collects splinters shrapnel and fragments.
  *Note.* The inflatable towers (masts) have sectionated stacked balloon design and small internal pressure. If some section is damaged, the leakage is easy compensated for by ground ventilators (increased fan flow)

  The other protection method for an entire city by AB-Net is shown in Fig.5b. The city is covered by light inflatable AB-Dome film as described in [1] - [14]. Below this film is located the AB-Nets 22 - 24 suspended by mounting fastening hardpoints on the dome cover, as it is shown in Fig. 5b. This design has advantages – it does not need support masts (instead hanging from the overpressure suspended overroof) and the climate inside the Dome is controlled [1]-[14].

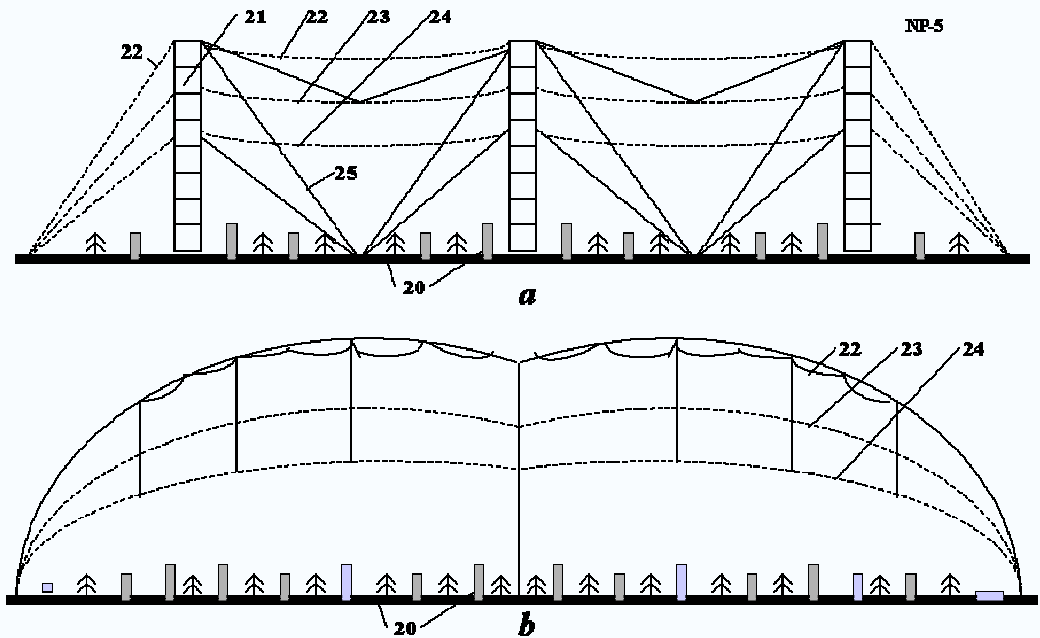

**Fig.5.** Protection by AB-Net of a city (base, big important object) from rockets and projectile. (*a*) Protection by nets suspended in the inflatable sectionated stacked balloon towers (masts) ; (*b*) Protection by nets suspended in inflatable dome. *Notations*: 20 – city (base or important object); 21 - inflatable sectionated stacked balloon towers (masts); 22 – parachute net; 23 – capture net; 24 – debris (fragments) net (if projectile will explode); 25 – bracing cable.

The methods of protection in different cases are shown in Fig. 6a-d. Figure 6a is protection of soldiers in the front line of battle. The inflatable (or steel) sectionated stacked balloon masts 3 are installed forward of our soldiers. The parachute net 2 is supported by masts. The AB-Net is designed thus (and this is part of the invention) that it joins the braking small parachutes to projectiles flying from the enemy side, but allows passage from the friendly side to the enemy side (the parachutes don't connect or not open). In sum, the enemy cannot kill the our soldiers, but our soldiers can repulse the attack. *Note*: Contrary to expectation, this protection is suitable for a fighting environment. The bullets and shells easily pass through the thin film of inflatable masts and produse only small holes equal in each case to the round's diameter. The air leakage throuth these holes is easily compensated for by small increases in the ventilator speed.

The figure 6b shows the protection of a vital road (highway) (or runway) from enemy sniping. The AB-Nets are installed along the dangerous part of road and protects the transport convoys (or expensive aircraft) from terrorist fire.

The AB-Net may be used in a counter-launch mode against small rockets and projectiles (Fig.6c). The locator system locates and computes the trajectory of incoming enemy projectiles. An anti-projectile cannon shoots a special shell in trajectory toward the enemy projectile. The anti-shell crosses the trajectory of the oncoming enemy shell. At the moment of contact, the shell snags the net and its' small brake parachute triggers. The parachute opens and brakes the enemy shell and does not allow the shell to reach its' aimpoint.

This system has considerable advantage in comparison to a more expensive rocket based APS. This system is cheaper because it doesn't need complex expensive self-guided missiles whose guidance system is destroyed after one use. The cannon doesn't need a precise pointing system. (The error margin is covered by the spreading net, not super-precise aiming, multiple rounds or a mid-course maneuver). The same system, though with a different net grid size, may be used against strategic rockets and their warheads having speed up 1500 m/s in end of trajectory (Fig.6d).

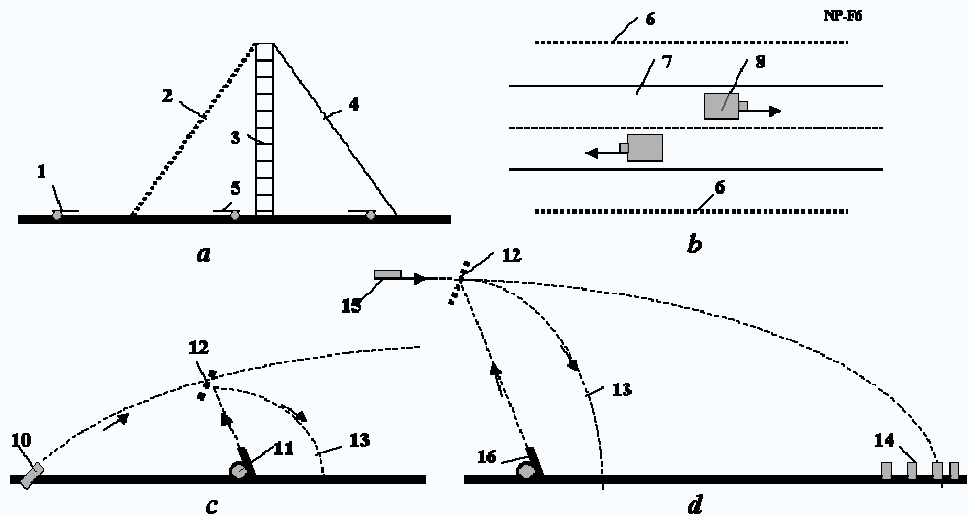

**Fig.6.** Protection from projectiles by AB-Nets: (***a***) Proiection of battle-front line; (***b***) Protection of the car (track) road; (***c***) Protection of object (for example, small town) from small (tactic) rockets; (***d***) Protection of a city from strategic missiles. *Notations*: 1 –enemy; 2 – parachute AB-Net; 3 – inflatable sectionated stacked balloon towers (or steel masts); 4 – bracing cable; 5 – our soldiers; 6 – ground parachute nets; 7 car (track) road; 8 – car, track; 10 – launch of enemy rocket; 11 – our cannon for shooting the parachute net; 12 – parachute air net; 13 – brake rocket trajectory; 14 - city; 15 - strategic missile or nuclear warhead; 16 – our cannon for shooting the parachute AB-Net.

## Theory and computation (estimation) of AB-Net

### a) General information

The efficiency of the offered net significantly depends from used artificial fiber.
 **1. Artificial fiber and cable properties** [15]-[20]**.** Cheap artificial fibers are currently being manufactured, which have tensile strengths of 3-5 times more than steel and densities 4-5 times less than steel. There are also experimental fibers (whiskers) that have tensile strengths 30-100 times more than steel and densities 2 to 5 times less than steel. For example, in the book [16] p.158 (1989), there is a fiber (whisker) $C_D$, which has a tensile strength of $\sigma = 8000$ kg/mm$^2$ and density (specific gravity) of $\gamma = 3.5$ g/cm$^3$. If we use an estimated strength of 3500 kg/mm$^2$ ($\sigma = 7 \cdot 10^{10}$ N/m$^2$, $\gamma = 3500$ kg/m$^3$), than the ratio is $\gamma/\sigma = 0.1 \times 10^{-6}$ or $\sigma/\gamma = 10 \times 10^6$. Although the described (1989) graphite fibers are strong ($\sigma/\gamma = 10 \times 10^6$), they are at least still ten times weaker than theory predicts. A steel fiber has a tensile strength of 5000 MPA (500 kg/sq.mm), the theoretical limit is 22,000 MPA (2200 kg/mm$^2$) (1987); polyethylene fiber has a tensile strength 20,000 MPA with a theoretical limit of 35,000 MPA (1987). The very high tensile strength is due to its nanotube structure [19].

Apart from unique electronic properties, the mechanical behavior of nanotubes also has provided interest because nanotubes are seen as the ultimate carbon fiber, which can be used as reinforcements in advanced composite technology. Early theoretical work and recent experiments on individual nanotubes (mostly MWNT's, Multi Wall Nano Tubes) have confirmed that nanotubes are one of the stiffest materials ever made. Whereas carbon-carbon covalent bonds are one of the strongest in nature, a structure based on a perfect arrangement of these bonds oriented along the axis of nanotubes would produce an exceedingly strong material. Traditional carbon fibers show high strength and stiffness, but fall far short of the theoretical, in-plane strength of graphite layers by an order of magnitude. Nanotubes come close to being the best fiber that can be made from graphite.

For example, whiskers of Carbon nanotube (CNT) material have a tensile strength of 200 Giga-Pascals and a Young's modulus over 1 Tera Pascals (1999). The theory predicts 1 Tera Pascals and a Young's modules of 1-5 Tera Pascals. The hollow structure of nanotubes makes them very light (the specific density varies from 0.8 g/cc for SWNT's (Single Wall Nano Tubes) up to 1.8 g/cc for MWNT's, compared to 2.26 g/cc for graphite or 7.8 g/cc for steel). Tensile strength of MWNT's nanotubes reaches 150 GPa.

Specific strength (strength/density) is important in the design of the systems presented in this paper; nanotubes have values at least 2 orders of magnitude greater than steel. Traditional carbon fibers have a specific strength 40 times that of steel. Since nanotubes are made of graphitic carbon, they have good resistance to chemical attack and have high thermal stability. Oxidation studies have shown that the onset of oxidation shifts by about $100^0$ C or higher in nanotubes compared to high modulus graphite fibers. In a vacuum, or reducing atmosphere, nanotube structures will be stable to any practical service temperature (in vacuum up 2800 $^oC$. in air up 750$^oC$).

In theory, metallic nanotubes can have an electric current density (along axis) more than 1,000 times greater than metals such as silver and copper. Nanotubes have excellent heat conductivity along axis up 6000 W/m·K. Copper, by contrast, has only 385 W/m·K.

About 60 tons/year of nanotubes are produced now (2007). Price is about $100 - 50,000/kg. Experts predict production of nanotubes on the order of 6000 tons/year and with a price of $1 – 100/kg to 2012.

Commercial artificial fibers are cheap and widely used in tires and countless other applications. The authors have found only older information about textile fiber for inflatable structures (Harris J.T., Advanced Material and Assembly Methods for Inflatable Structures, AIAA, Paper No. 73-448, 1973). This refers to DuPont textile Fiber **B** and Fiber **PRD-49** for tire cord. They are 6 times strong as steel (psi is 400,000 or 312 kg/mm$^2$) with a specific gravity of only 1.5. Minimum available yarn size (denier) is 200, tensile module is $8.8 \times 10^6$ (**B**) and $20 \times 10^6$ (**PRD-49**), and ultimate elongation (percent) is 4 (**B**) and 1.9 (**PRD-49**). Some data are in Table 1.

**Table 1.** Material properties

| Material | Tensile strength kg/mm$^2$ | Density g/cm$^3$ | **Fibers** | Tensile strength kg/mm$^2$ | Density g/cm$^3$ |
|---|---|---|---|---|---|
| Whiskers | | | | | |
| AlB$_{12}$ | 2650 | 2.6 | QC-8805 | 620 | 1.95 |
| B | 2500 | 2.3 | TM9 | 600 | 1.79 |
| B$_4$C | 2800 | 2.5 | Allien 1 | 580 | 1.56 |
| TiB$_2$ | 3370 | 4.5 | Allien 2 | 300 | 0.97 |
| SiC | 1380-4140 | 3.22 | Kevlar or Twaron | 362 | 1.44 |
| **Material** | | | Dynecta or Spectra | 230-350 | 0.97 |
| Steel prestressing strands | 186 | 7.8 | Vectran | 283-334 | 0.97 |
| Steel Piano wire | 220-248 | | E-Glass | 347 | 2.57 |
| Steel A514 | 76 | 7.8 | S-Glass | 471 | 2.48 |
| Aluminum alloy | 45.5 | 2.7 | Basalt fiber | 484 | 2.7 |
| Titanium alloy | 90 | 4.51 | Carbon fiber | 565 | 1,75 |
| Polypropylene | 2-8 | 0.91 | Carbon nanotubes | 6200 | 1.34 |

Source: [15]-[18]. Howatsom A.N., Engineering Tables and Data, p.41.

Industrial fibers have up to $\sigma = 500\text{-}600$ kg/mm$^2$, $\gamma = 1500\text{ -}1800$ kg/m$^3$, and $\sigma/\gamma = 2,78 \times 10^6$. But we are projecting use in the present projects the cheapest films and cables applicable (safety $\sigma = 100 - 200$ kg/mm$^2$).

**2. Leakage of air through inflatable tower (mast) hole.** The leakage of air through bullet holes, requested power of fans (ventilator), and time of sinking of tower (mast) cover (in case of large hole) may be estimated by the equation:

$$V = \sqrt{\frac{2p}{\rho}}, \quad M_a = \rho V S_h, \quad N = \frac{pVS_h}{\eta}, \qquad (1)$$

where $V$ is speed of air leakage, m/s; $p$ is overpressure, N/m$^2$; $\rho$ is air density at given altitude, $\rho$ = 1,225 kg/m$^3$ at $H$ = 0; $S_h$ is area of hole, m; $N$ is motor power, W; $\eta$ is coefficient efficiency of ventilator.

*Example*. The total area of holes equals $S_h$ = 1 m$^2$ (10×10 m) at $H$ = 0 m, $p$ = 0.001 atmosphere = 100 N/m$^2$, $\eta$ = 0.8. Computation gives the $V$ = 12.8 m/s, $M_a$ = 15.7 kg/s, $N$ =1.6 kW.

**3. The maximum specific wind dynamic pressure** to inflatable mast (tower) from wind is:

$$p_w = \frac{\rho V^2}{2}, \qquad (2)$$

where $\rho$ = 1.225 kg/m$^3$ is air density; $V$ is wind speed, m/s; $p_w$ is specific dynamic wind pressure to 1 m$^2$, N/m$^2$.

For *example*, a storm wind with speed $V$ = 20 m/s, standard air density is $\rho$ = 1.225 kg/m$^3$. Than dynamic pressure is $p_w$ = 245 N/m$^2$. That is four times less when internal pressure $p$ = 1000 N/m$^2$ = 0,01 atm. When the need arises, sometimes the internal pressure can be voluntarily decreased, bled off.

## b) Method of Computation and Estimation

**1. Brake projectile distance having brake parachute.** The motion of a projectile with parachute is described by differential equations:

$$\frac{dV}{dt} = -C_d \frac{\rho S}{2m} V^2 = -aV^2, \quad \frac{dL}{dt} = V, \quad \text{where} \quad a = C_d \frac{\rho S}{2m} \approx 0.6 \frac{S}{m}, \qquad (3)$$

where $V$ is projectile speed, m/s, $V \geq V_{min} = \sqrt{g/a}$; $t$ is time, sec; $C_d \approx 1$ is parachute drag coefficient; $\rho$ = 1.225 kg/m$^3$ is standard air density; $S$ is projection of parachute area, m$^2$; $m$ is mass of system (projectile + parasite + connection cable (part of AB-subnet)), kg; $L$ is distance of fly projectile, m; $a$ is drag constant, 1/m.

After integration of equations (3) we get

$$L = -\frac{1}{a} \ln \frac{V}{V_0} \quad \text{or} \quad V = V_0 e^{-aL}, \qquad (4)$$

where $V_0$ is initial speed of projectile, m/s.

The computation of equation (4) is presented in Fig. 7.

**2. The maximal braking force and overload of projectile** compute by equations:

$$P_m = amV_0^2, \quad n_m = \frac{P_m}{mg} = \frac{a}{g} V_0^2, \qquad (5)$$

where $P_m$ is maximal braking force of projectile (in moment of connection to parachute), N; $n_m$ is maximal overload of projectile, in "g"; $g$ = 9.81 m/s$^2$ is Earth's gravitation.

The computation is shown in Fig.8.

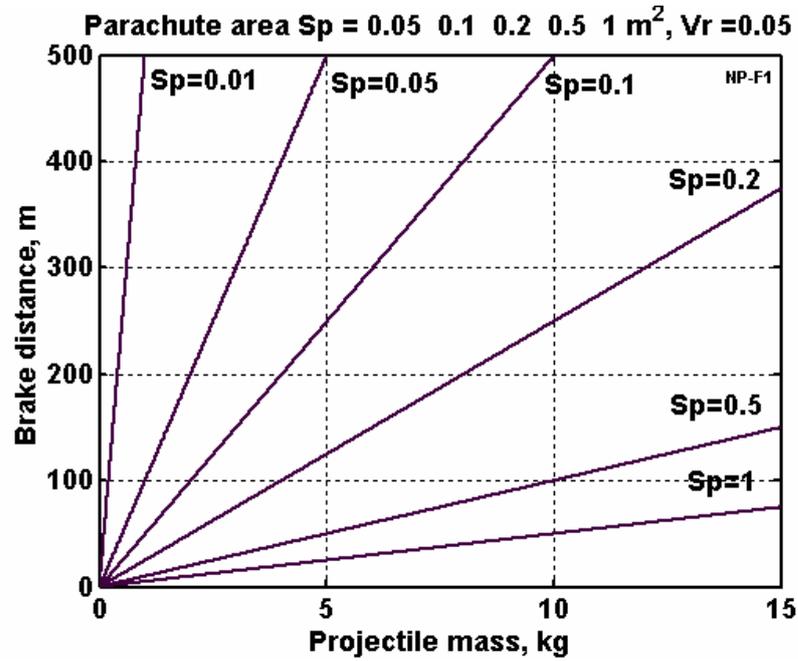

**Fig. 7.** The brake distance of projectile versus the projectile mass (0 ÷ 15 kg) and area of brake parachute $S = S_p = (0.05 \div 1\ m^2)$ for relative finish projectile speed $V_r = (V/V_0) = 0.05$.

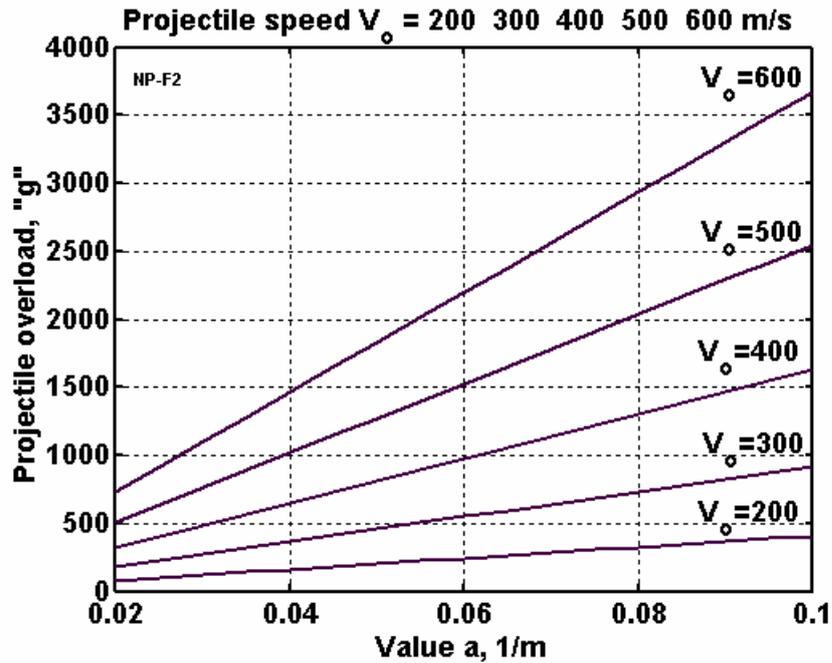

**Fig. 8.** Maximal projectile overload (in "g") via projectile speed $V_0$ and coefficient $a$.

3. **The cross-section and diameter of subnet cable** compute by equation:

$$s = \frac{P_m}{2\sigma} \approx 5\frac{mn_m}{\sigma},\quad d = \sqrt{\frac{4s}{\pi}}, \qquad (6)$$

where $s$ is cross-section of subnet cables, m$^2$; $\sigma$ is safety tensile stress of subnet cable, N/m$^2$; $d$ is diameter of subnet cable, m. We assume that projectile tears (loads) ONLY TWO subnet cables. (A conservative estimate).

The computation is presented in Fig.9.

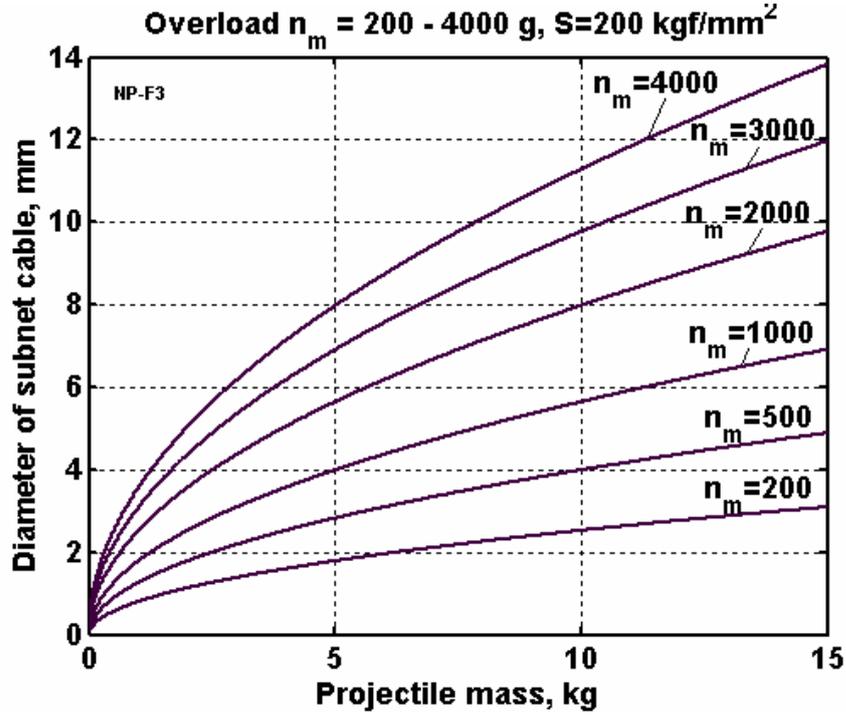

**Fig.9.** Diameter of subnet cable (mm) versus mass of projectile and maximal overload.

4. **1 m² mass of parachute net and mass of subnet** may be estimated by the equations:
   a) Mass of parachute

$$p = \frac{\rho V_0^2}{2} \approx 0.61 V_0^2, \quad r = \sqrt{\frac{S}{\pi}}, \quad \delta = \frac{pr}{2\sigma_p}, \quad m_1 = 2\gamma \delta S, \tag{8}$$

where $m_1$ is mass of parachute, kg; $\gamma \approx 1500$ kg/m³ is specific density of parachute material, $S$ is projection of parachute area, m²; $\delta$ is thickness of parachute material, m; $p$ is air dynamic pressure, N/m²; $r$ is radius of parachute, m; $\sigma_p$ is safety stress of parachute, N/m². $\sigma_p \approx 100$ kgf/mm² = $10^9$ N/m².

   b) Mass of subnet cables:

$$m_2 = 2\gamma s_1/l_1 . \tag{7}$$

where $m_2$ is mass of subnet cable, kg/m²; $\gamma \approx 1500$ kg/m³ is specific density of cable material, kg/m³; $s_1$ cross-section of subnet cable, m²; $l_1$ is step of subnet cables, m.

   1 m² of parachute net for step of main net $l_2$ and cell area $S_b = l_2^2$ requests $N = 1/S_b$ parachutes/m² .

   c) Mass of main net cables is computed as:

$$m_3 = 2\gamma s_2/l_2 \tag{9}$$

where $m_3$ is mass of net main cable, kg/m²; $\gamma \approx 1500$ kg/m³ is specific density of cable material, kg/m³; $s_2 \approx (2 \div 3)s_1$ cross-section of net main cable, m²; $l_2$ is mesh size (step size) of big net (main cables), m.

   The total 1 m² mass of parachute net

$$m_n = Nm_1 + m_2 + m_3 , \tag{10}$$

Mass of one cell subnet cable and one parachute:

$$m_s = m_1 + m_2 S_b . \tag{11}$$

The result of computations yielding the mass of net per 1 m² is presented in Fig.10.

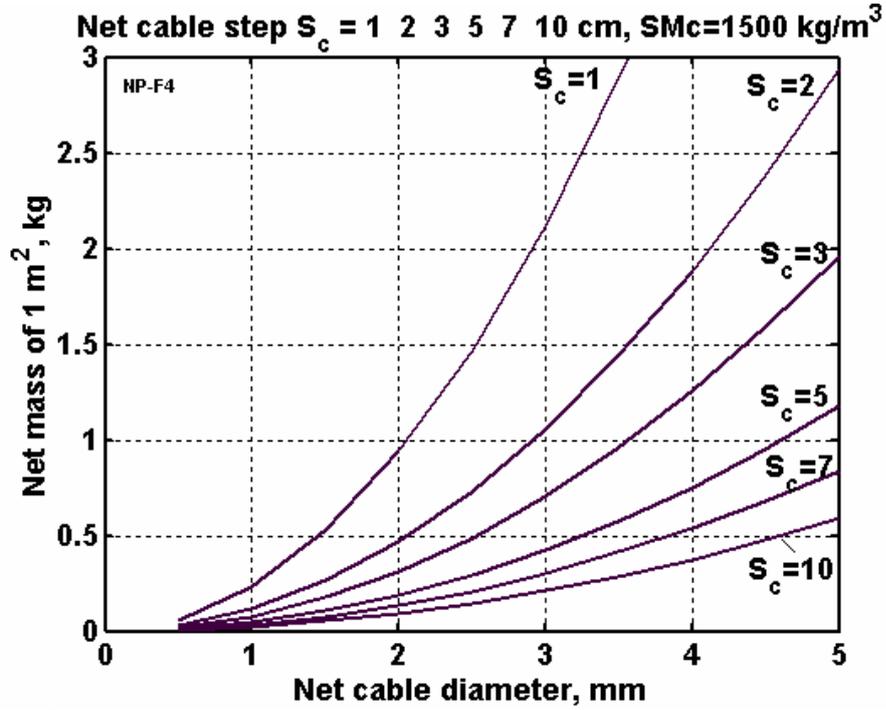

**Fig.10**. One m² net mass via diameter of the net cable (0.5 - 5 mm) and net mesh size (cable stepping distance) (1 – 10 cm).

5. **Tear out the subnet and parachute by projectile from AB-Net.** The overload and force in impact moment may be estimated by formulas

$$n_s = \frac{V_0^2}{2gL_2}, \quad P_s = gn_s m_s, \quad \text{from} \quad P_s = P_m \quad \text{we get} \quad L_2 = \frac{m_s}{am} \approx \frac{m_s}{0.6S}, \qquad (12)$$

where $L_2$ is tear out distance, m ($L_2 \approx (1 \div 2)\, l_2$); $n_s$ is tear out overload, "g"; $m_s$ is mass of one cell subnet cable + parachute, kg (Eq.(11)). We assume that projectile tears ONLY TWO subnet cables.

6. **Cost of 1 m² net**. The cost of 1 m² net approximately is
$$C = cm_n,$$
where $C$ is cost of 1 m² net, $/m²; $c$ is cost 1 kg artificial fiber, $/kg; $m_n$ is mass of 1 m² of net, kg/m² (Eq.(10)).

**Computation of lower (capture) net**

The purpose of this capture net is to catch (collect) the projectiles of sufficiently low speed that capture can be accomplished without damage to the net. This net is needed in cases when the main (parachute dispensing slowdown net) is located such (for example, horizontal (as on a roof) and the already braked projectile must not reach the city after extinction of momentum because the round is still live and can have salvage fuzing with, for example, an ignition (detonation) delay.

The flexure of the capture net, maximal force, and projectile overload may be estimated by formulas

$$\text{From} \quad \frac{mV^2}{2} = k\frac{l_c^2}{2} \quad \text{we get} \quad l_c = \sqrt{\frac{m}{k}}V, \quad P = \sqrt{km}\,V, \quad n_c = \frac{P}{mg}, \qquad (13)$$

where $l_c$ is flexure of the capture net, m; $k$ is net elasticity, N/m; $m$ is projectile mass, kg; $V$ is **braked** projectile speed, m/s; $P$ is maximal net force, N; $n_s$ is projectile overload, "g".

The cross-section area of net cable $s_c$, diameter of net cable $d_c$, tensile net stress $F$, and 1 m² cable mass $m_c$ may be estimate the following equations (in metric system):

$$s_c = \frac{P}{2\sigma}, \quad d_c = \sqrt{\frac{4s_c}{\pi}}, \quad F \approx \frac{PL_c}{16l_c}, \quad m_c = 2\gamma\frac{s_c}{l_c}, \quad (14)$$

where $L_c$ is approximately distance between the inflatable masts (towers), m. $l_c$ is cable step in capture net

We assume that projectile tears ONLY TWO subnet cables everywhere. This is the conservative case; more will be more effective in stopping the projectile.

## Projects

**Project #1. Protection city from Qassam (Kassam-3) rockets** (Fig.5a).

Let us to consider the protection of a typical small city from Kassam-3 rockets having diameter $D = 17$ cm, flight mass (after launch) $m = 30$ kg, and landing speed $V_0 = 300$ m/s (in reality $V_0 \approx 250$ m/s). Assume the city has AB-Net roof, located on inflatable masts having radius $R = 10$ m and altitude 450 m. We calculate the data and cost for area 1 km² and 10 km². If you know the city area, you easily can recalculate the data for the new area.

**1. Computation of parachute net.** Let us take $S = 0.5$ m² and $V/V_0 = 0.1$, $V = 30$ m/s ≈ $V_{min} = 31.3$ m/s. From equations (4)-(14) we find:
1) Coefficient $a = 0.6S/m = 0.6 \times 0.5/30 = 0.01$ [1/m].
2) Braking distance $L = (-1/a) \times \ln(V/V_0) = (-1/0.02) \times \ln(0.1) = 230$ m.
3) The maximal braking force and overload of projectile is computed by the equation:

$$P_m = amV_0^2 = 0.01 \cdot 30 \cdot 300^2 = 2.7 \cdot 10^4 \ N, \quad n_m = \frac{P_m}{mg} = \frac{2.7 \cdot 10^4}{30 \cdot 9.81} = 90 \ g \ .$$

4) The cross-section and diameter of subnet cable are computed by the equation (for $\sigma = 200$ kg/mm² = $2 \times 10^9$ N/m²):

$$s_1 = \frac{P_m}{2\sigma} = \frac{2.7 \cdot 10^4}{2 \cdot 2 \cdot 10^9} = 6.75 \cdot 10^{-6} \ m^2 = 6.75 \ mm^2, \quad d = \sqrt{\frac{4s_1}{\pi}} = 2.93 \ mm \ .$$

5) 1 m² mass of subnet cables, mass of parachute, and cell mass of subnet, 1 m² mass of total parachute net (for $\gamma = 1500$ kg/m², $l_1 = 0.1$ m) are:
  a) Mass of one parachute for safety parachute stress $\sigma_p = 10^9$ N/m²:

$$p = \frac{\rho V_0^2}{2} = 0.61 \cdot 300^2 = 5.51 \cdot 10^4 \ \frac{N}{m^2}, \quad r = \sqrt{\frac{S}{\pi}} = 0.4 \ m, \quad \delta = \frac{pr}{2\sigma_p} = 1.1 \cdot 10^{-5} \ m,$$

$$m_1 = 2\gamma\delta S = 2 \cdot 1500 \cdot 1.1 \cdot 10^{-5} \cdot 0.5 = 0.0165 \ kg$$

1 m² of parachute net for step of main net $l_2 = 0.5$ m and cell area $S_b = l_2^2 = 0.25$ m² requests $N = 1/S_b = 4$ parachutes/m².
  b) 1 m² mass of subnet cables having small net step $l_1 = 0.1$ m:

$$m_2 = \frac{2\gamma s_1}{l_1} = \frac{2 \cdot 1500 \cdot 6.75 \cdot 10^{-6}}{0.1} = 0.2 \ kg/m^2 \ .$$

  c) Mass of 1 m² main net cables (for $s_2 = 2s_1$, $S_b = 0.25$ m²) is:

$$m_3 = \frac{2\gamma s_2}{l_2} = \frac{2 \cdot 1500 \cdot 2 \cdot 6.75 \cdot 10^{-6}}{0.5} = 0.081 \ kg/m^2 \ .$$

The total 1 m² mass of parachute net
$m_n = Nm_1 + m_2 + m_3 = 4 \times 0.0165 + 0.2 + 0.081 = 0.347$ kg/m²,
Mass of one cell subnet cable and one parachute ($S_b = 0.25$ m²):
$m_s = m_1 + m_2 S_b = 0.0165 + 0.2 \times 0.25 = 0.0665$ kg.

6) Parameters of tearing out the subnet and parachute by projectile from AB-Net: The overload and force in

impact moment is ($L_2 = (1 \div 2) \times l_2$):

$$n_s = \frac{V_0^2}{2gL_2} = \frac{300^2}{2 \cdot 9.81 \cdot 1} = 4.6 \cdot 10^3 \ g,$$

$$P_s = gn_s m_s = 9.81 \cdot 4.6 \cdot 10^3 \cdot 0.0665 = 0,3 \cdot 10^4 \ N < P_m = 2.7 \cdot 10^4 \ N$$

The difference between flight braking force $P_m$ and tear out force $P_s$ is very high. That means the parachute area is large; our design is far from optimum. But we cannot decrease the parachute area because the braking distance $L = 230$ m (the distance between parachute and capture nets) requires that much braking power

In any case the AB-Net requires testing for tuning the cross-section cable area and spacing of the net mesh for any given projectile. These tests are cheap.

**2. Computation of capture net.** Let us take the net elasticity $k = 500$ N/m and final braking speed of Qassam rocket $V = 30$ m/s (see above), $m = 30$ kg.

The flexure of capture net $l_c$, maximal force $P$, and projectile overload $n_c$ may be estimated by formulas (Eq. (13)):

$$l_c = \sqrt{\frac{m}{k}} V = \sqrt{\frac{30}{500}} 30 = 7.35 \ m, \quad P = \sqrt{km} V = \sqrt{500 \cdot 30} \cdot 30 = 3674 \ N, \quad n_c = \frac{P}{mg} = 12.5 \ g,$$

where $l_c$ is maximal flexure of the capture net, m; $k$ is net elasticity, we take $k = 500$ N/m; $m$ is projectile mass, kg; $V$ is **braked** projectile speed, m/s; $P$ is maximal net force, N; $n_s$ is projectile overload, "g".

Let us take the distance between the support cable $L_c = 100$ m, the safety tensile stress $\sigma = 2 \times 10^9$ N/m². The cross-section area of net cable $s_c$, diameter of net cable $d_c$, tensile net stress $F$, and 1 m² cable mass $m_c$ may be estimate the following equations (in metric system) (Eq. (14)):

$$s_c = \frac{P}{2\sigma} = \frac{3674}{2 \cdot 2 \cdot 10^9} = 0.92 \cdot 10^{-6} \ m^2, \quad d_c = \sqrt{\frac{4s_c}{\pi}} = 1.08 \ mm,$$

$$F \approx \frac{PL_c}{16 l_c} = \frac{3674 \cdot 100}{16 \cdot 7.35} = 3124 \ N, \quad m_c = 2\gamma \frac{s_c}{l_c} = 2 \cdot 1500 \cdot \frac{0.92 \cdot 10^{-6}}{0.1} = 0.0276 \ kg/m^2$$

The total mass of nets is
$m_t = m_n + m_c = 0.347 + 0.028 = 0.375$ kg/m².

We assume, as before, that the projectile tears ONLY TWO subnet cables. This is the conservative case; more will be more effective in stopping the projectile.

**3. Cost of net and protection**. The price of the glass fiber is 0.7$/kg. If price of artificial fiber is $c = 2$/kg, the cost of 1 m² net is about $C = 0.75$/m². The net for 1 km² costs about $0.75 M/km², for small city having area 10 km² the net cost $7.5 million. We need 1- 2 masts in 1 km². If the inflatable support mast has diameter 20 m, height 500 m and cost about $10,000 each, the total mast cost is about $200,000. The protection from Qassam rockets of the small city will costs about $8 -10 millions.

### Project #2. Protection of battle-front from shells

having mass $m = 6.4$ kg, diameter $D = 76$ mm and speed $V_0 = 600$ m/s, Fig.4a.

Let us to consider the protection of a battle-front from a modern typical 76 mm caliber cannon having 6,4 kg shell. The nozzle speed of the shell is about 600 ÷750 m/s, but at distance ~200-400m m the air drag decreases this speed to 600 m/s. That way we take in our computation the contact speed $V_0 = 600$ m/s.

We calculate the data and cost for front line 1 km. If you know the front length, you easy recalculate the data for new length.

**1. Computation of parachute net.** Let us take the parachute area $S = 0.15$ m$^2$ and $V/V_0 = 0.05$, $V = 30$ m/s. The $V_{min} = 26.7$ m/s. From equations (4)-(14) we find:
1) Coefficient $a = 0.6S/m = 0.6 \times 0.15/6.4 = 0.014$ [1/m].
2) Braking distance $L = (-1/a) \times \ln(V/V_0) = (-1/0,014) \times \ln(0.05) = 214$ m.
3) The maximal braking force and overload of projectile compute by equations:

$$P_m = amV_0^2 = 0{,}014 \cdot 6.4 \cdot 600^2 = 3.22 \cdot 10^4 \ N, \quad n_m = \frac{P_m}{mg} = \frac{3.22 \cdot 10^4}{6.4 \cdot 9.81} = 513 \ g \ .$$

4) The cross-section and diameter of subnet cable are computed by equation (for $\sigma = 200$ kg/mm$^2$ $= 2 \times 10^9$ N/m$^2$):

$$s_1 = \frac{P_m}{2\sigma} = \frac{3.22 \cdot 10^4}{2 \cdot 2 \cdot 10^9} = 8 \cdot 10^{-6} \ m^2 = 8 \ mm^2, \quad d = \sqrt{\frac{4s}{\pi}} = 3.2 \ mm \ .$$

5) 1 m$^2$ mass of subnet cables, mass of parachute, and cell mass of subnet, 1 m$^2$ mass of total parachute net (for $\gamma = 1500$ kg/m$^2$, $l_1 = 0.04$ m, $l_2 = 0.2$ m) are:

 a) Mass of one parachute for safety parachute stress $\sigma_p = 10^9$ N/m$^2$ :

$$p = \frac{\rho V_0^2}{2} = 0.61 \cdot 600^2 = 22 \cdot 10^4 \ \frac{N}{m^2}, \quad r = \sqrt{\frac{S}{\pi}} = 0.22 \ m, \quad \delta = \frac{pr}{2\sigma_p} = 2.42 \cdot 10^{-5} \ m,$$

$$m_1 = 2\gamma\delta S = 2 \cdot 1500 \cdot 2.42 \cdot 10^{-5} \cdot 0.15 = 0.011 \ kg$$

1 m$^2$ of parachute net for step of main net $l_2 = 0.2$ m and cell area $S_b = l_2^2 = 0.04$ m$^2$ requests $N = 1/S_b = 25$ parachutes/m$^2$.

 b) 1 m$^2$ mass of subnet cables having small net step $l_1 = 0.04$ m:

$$m_2 = \frac{2\gamma s_1}{l_1} = \frac{2 \cdot 1500 \cdot 8 \cdot 10^{-6}}{0.04} = 0.6 \ kg/m^2 \ .$$

 c) Mass of 1 m$^2$ main net cables (for $s_2 = 2s_1$, $S_b = 0.04$ m$^2$) is:

$$m_3 = \frac{2\gamma s_2}{l_2} = \frac{2 \cdot 1500 \cdot 2 \cdot 8 \cdot 10^{-6}}{0.2} = 0.24 \ kg/m^2 \ .$$

 The total 1 m$^2$ mass of parachute net
 $m_n = Nm_1 + m_2 + m_3 = 25 \times 0.011 + 0.6 + 0.24 = 0.851$ kg ,
 Mass of one cell subnet cable and one parachute ($S_b = 0.04$ m$^2$):
 $m_s = m_1 + m_2 S_b = 0.011 + 0.6 \times 0.04 = 0.035$ kg.

6) Tear out the subnet and parachute by projectile from AB-Net. The overload and force in impact moment is ($L_2 = (1 \div 2) \times l_2$):

$$n_s = \frac{V_0^2}{2gL_2} = \frac{600^2}{2 \cdot 9.81 \cdot 0.4} = 4.5 \cdot 10^4 \ g,$$

$$P_s = gn_s m_s = 9.81 \cdot 4.5 \cdot 10^4 \cdot 0.035 = 1.54 \cdot 10^4 \ N < P_m = 3.22 \cdot 10^4 \ N$$

Our net mass is high. We can decrease it up $m_n = 0.5$ kg/m$^2$ if we decrease the parachute area up $S = 0,1$ m$^2$. But braking distance increases up $L = 300$ m.

We assume that projectile tears ONLY TWO subnet cables.

**3. Cost of net and protection**. If price of artificial fiber is $c = 2$\$/kg, the cost of 1 m$^2$ net is about $C = 1.8$\$/m$^2$. The net for $1000 \times 20$ m$^2$ costs about \$36,000/km. If the support mast has height 20 m and cost about \$100 each, and distance between them is 50 m, the total mast cost is about \$2,000/km. The protection from cannon shells of the 1 km of battle-front will costs about \$40,000/km.

# Project #3. Protection of place, road, base, meeting, building, etc. from Kalashnikov sub-machine gun.

The Kalashnikov sub-machine gun is widely used by terrorists. Caliber is 7.6 mm, the nozzle speed is 715 m/s but at distance 150-250 m the bullet speed is about $V_0 = 600$ m/s,. Assume the mass of bullet is m = 10 g = 0.01 kg.

Let us to consider the protection of a battle-front from the Kalashnikov sub-machine gun. We calculate the data and cost for front line 1 km. If you know the battlefront's length, you easily may recalculate the data for the given installation.

**1. Computation of parachute net.** Let us take $S = 10$ cm$^2 = 10^{-3}$ m$^2$ and $V/V_0 = 0.05$. From equations (4)-(14) we find:
1) Coefficient $a = 0.6S/m = 0.6 \times 10^{-3}/10^{-2} = 0.06$ [1/m], $V_{min} = 13$ m/s.
2) Braking distance $L = (-1/a) \times \ln(V/V_0) = (-1/0,06) \times \ln(0.05) = 50$ m.
3) The maximal braking force and overload of projectile compute by equations:

$$P_m = amV_0^2 = 0,06 \cdot 0.01 \cdot 600^2 = 2.16 \cdot 10^2 \text{ N}, \quad n_m = \frac{P_m}{mg} = \frac{2.16 \cdot 10^2}{10^{-2} \cdot 9.81} = 2.16 \cdot 10^3 \text{ g}.$$

4) The cross-section and diameter of subnet cable are computed by equation (for $\sigma = 100$ kg/mm$^2 = 2 \times 10^9$ N/m$^2$):

$$s_1 = \frac{P_m}{2\sigma} = \frac{2.16 \cdot 10^2}{2 \cdot 10^9} = 1.08 \cdot 10^{-7} \text{ m}^2 = 0.108 \text{ mm}^2, \quad d = \sqrt{\frac{4s}{\pi}} = 0.14 \text{ mm}.$$

5) 1 m$^2$ mass of subnet cables, mass of parachute, and cell mass of subnet, 1 m$^2$ mass of total parachute net (for $\gamma = 1500$ kg/m$^2$, $l_1 = 0.004$ m, $l_2 = 0.03$ m) are:

 a) Mass of one parachute for safety parachute stress $\sigma_p = 10^9$ N/m$^2$:

$$p = \frac{\rho V_0^2}{2} = 0.61 \cdot 600^2 = 22 \cdot 10^4 \frac{\text{N}}{\text{m}^2}, \quad r = \sqrt{\frac{S}{\pi}} = 0.018 \text{ m}, \quad \delta = \frac{pr}{2\sigma_p} = 2 \cdot 10^{-6} \text{ m},$$

$$m_1 = 2\gamma\delta S = 2 \cdot 1500 \cdot 2 \cdot 10^{-6} \cdot 0.001 = 6 \cdot 10^{-6} \text{ kg}$$

1 m$^2$ of parachute net for step of main net $l_2 = 0.03$ m and cell area $S_b = l_2^2 = 0.0009$ m$^2$ requests $N = 1/S_b = 1111$ parachutes/m$^2$.

 b) 1 m$^2$ mass of subnet cables having small net step $l_1 = 0.004$ m:

$$m_2 = \frac{2\gamma s_1}{l_1} = \frac{2 \cdot 1500 \cdot 1.08 \cdot 10^{-7}}{0.004} = 0.081 \text{ kg/m}^2.$$

 c) Mass of 1 m$^2$ main net cables (for $s_2 = 2s_1$, $l_2 = 0.03$ m$^2$) is:

$$m_3 = \frac{2\gamma s_2}{l_2} = \frac{2 \cdot 1500 \cdot 2 \cdot 1.08 \cdot 10^{-7}}{0.03} = 0.022 \text{ kg/m}^2.$$

The total 1 m$^2$ mass of parachute net
$m_n = Nm_1 + m_2 + m_3 = 1111 \times 6 \times 10^{-6} + 0.081 + 0.022 = 0.11$ kg,
Mass of one cell subnet cable and one parachute ($S_b = 0.0009$ m$^2$):
$m_s = m_1 + m_2 S_b = 6 \times 10^{-6} + 0.081 \times 0.0009 = 78 \times 10^{-6}$ kg.

6) Tear out the subnet and parachute by projectile from AB-Net. The overload and force in impact moment is ($L_2 = (1 \div 2) \times l_2$):

$$n_s = \frac{V_0^2}{2gL_2} = \frac{600^2}{2 \cdot 9.81 \cdot 0.06} = 30 \cdot 10^4 \text{ g},$$

$$P_s = gn_s m_s = 9.81 \cdot 30 \cdot 10^4 \cdot 78 \cdot 10^{-6} = 230 \text{ N} > P_m = 216 \text{ N}$$

We receive $P_s > P_m$. That means the cell size 3×3 cm is small. One must decreases to 4×4 cm. We assume that projectile tears ONLY TWO subnet cables.

**3. Cost of net and protection.** If price of artificial fiber is c = 3$/kg, the cost of 1 m$^2$ net is about C = 0.4$/m2. The net for 1000×10 m$^2$ costs about $4000/km and weighs <1350 kg. If the support mast has height of 10 m and cost about $50 each, and the distance between them is 20 m, the total mast cost is about $2500/km and each mast supports ~27 kg. The total protection from submachine gun and rifle of the 1 km of battle-front will cost about $6500/km (without labor cost).

The author is prepared to discuss the problems with organizations which are interested in research and development related projects.

## Discussion

For over a century has raged the competition between armor and shell. The tank armor has reached 200 mm thickness, the military ship armor 400 mm and more. The tank itself as a consequence became very expensive, heavy and awkward. The military ships became heavy, expensive and required deep water.

Entire generations have become accustomed to the idea of steel or superhard ceramic armor and when they hear the possibility raised that a thin net can protect from shells or bullets they may be forgiven for feeling a certain amount of disorientation.

But the author doesn't offer the thin net for a tank or ship. He offers protection instead for large territorries, cities, bases, roads,, battle-fronts—the entire range of what are now known as "soft targets" from enemy (terrorists) attacks – with the requirement only that they be places which can be separated by sufficient (projectile braking) distance from enemy fairing. He offers the interesting possibility of using the atmosphere (and the property of high air drag) as armor.

There are problems in adopting the idea to reality: how to join the rocket, shell or bullet to the small parachute or piece of net. But even supposing the case where, for example, the projectile did not successfully snag onto the parachute, the projectiles' very impact against the net creates a short high overload, changing the projectile trajectory; and the projectile misses its aimpoint.

The net is very cheap and the idea may be easily tested. The need, optimal parameters may be found by experiments.

The cost of the offered passive AB-Net Protection System may be less by hundred of times, then the cost of an *effective* active anti-rocket system (tens of billions $US). The anti-rocket system is useless in peacetime and it may be useless soon in wartime because of easily adopted enemy countermeasures.

The offered AB-Nets are different for different projectiles. The net against Kalashnikov automatic gunfire probably cannot join enough parachutes to a 76 mm gun shell. But in any case the impact to the net changes the projectile trajectory of projectile and one shot would miss its aim. A barrage, of course, as always, presents the target with greater dangers. For protection from different projectiles we can use the suitable nets located in ascending-size series.

The one lack of conventional AB-Nets –would be that they protect enemy from reciprocal fire. But the author has also designed a specialized AB-Net which protect one sides and still allows firing at the other side. He has numerous inventions concerning details of design: How the parachutes open, net form details, connection cables and the right way to align them, etc.

The associated problems are researched in author references [1]-[14].

## Results

Author offers the described inexpensive AB-Nets which protects cities, military bases, battle-front from rockets, mortar shell, projectiles, weapon (bombs) delivered by rockets, missiles, guns and aviation.

The principal advantages of the offered AB-Nets follow:

1. AB-Nets may be cheaper by hundreds to thousands of times than current comparable anti-rocket systems or defense systems.
2. The city employing the film AB-Dome for supporting the AB-Nets can has fine climate inside (the domed city)[1]. (this also may enable its financing).
3. These AB-Nets may be installed in some hours (the current defense zone against incoming supersonic weapons is built over some years and cost some billions of dollars).
4. AB-Nets do not need high technology and can be built by a poor country.
5. The AB-Net is easy tested and evaluated at low cost.
6. Author has a subdesign of AB-Nets which allow protecting our soldiers from enemy bullets and shells and simultaneously allow firing (shooting) towards the enemy through AB-Nets.
7. Height (up to 400-500 m) city AB-Nets may be used for high antennae site for regional TV, for cell phone communication, for long distance location (differential GPS service). (This also may enable its financing).
8. Unlike any known active ballistic missile defense, offered AB-Nets also can defend against submunitions and dispersed killing agents such as shrapnel.
9. The height over the city of suspended AB-Nets may be used for the high altitude windmills (and the getting of cheap renewable wind energy).
10. An AB-Net covered by thin film may be used for night illumination and entertainment or advertising (this also may enable its financing).

Additional applications of the closed author ideas the reader may find in [1]-[14].

## Acknowledgement

The author wishes to acknowledge Joseph Friedlander of Shave Shomron, Israel for correcting the author's English and useful technical advice and suggestions.